\documentclass[english,3p]{revtex4}
\usepackage[T1]{fontenc}
\usepackage[latin9]{inputenc}
\usepackage{amsmath}
\usepackage{amssymb, enumerate}
\usepackage{color}
\usepackage{babel}
\usepackage{graphicx}
\usepackage{epstopdf}
\usepackage{float}
\usepackage{thmtools}
\usepackage{thm-restate}
\usepackage{hyperref}
\usepackage{cleveref}
\makeatletter

\@ifundefined{textcolor}{}
{%
 \definecolor{BLACK}{gray}{0}
 \definecolor{WHITE}{gray}{1}
 \definecolor{RED}{rgb}{1,0,0}
 \definecolor{GREEN}{rgb}{0,1,0}
 \definecolor{BLUE}{rgb}{0,0,1}
 \definecolor{CYAN}{cmyk}{1,0,0,0}
 \definecolor{MAGENTA}{cmyk}{0,1,0,0}
 \definecolor{YELLOW}{cmyk}{0,0,1,0}
}

\newtheorem{theorem}{Theorem}
\newtheorem{corollary}{Corollary}
\newtheorem{proposition}{Proposition}

\newtheorem{lem}{Lemma}
\newtheorem{definition}{Definition}

\newcommand{\un}{1\mkern -4mu{\rm l}}

\newcommand{\ra}{\rangle}
\newcommand{\lar}{\longleftarrow}
\newcommand{\rar}{\longrightarrow}

\DeclareMathOperator{\Tr}{Tr}
\DeclareMathOperator{\tr}{tr}
\makeatother

\begin{document}

\title{Random and free observables saturate the Tsirelson bound for CHSH inequality}

\author{Z. Yin$^{1,2}$, A. W. Harrow$^{3}$, M. Horodecki$^{1}$, M. Marciniak$^{1}$ and A. Rutkowski$^{1}$}

\affiliation{$^1$
Faculty of Mathematics, Physics and Informatics, University of Gda\'{n}sk, 80-952 Gda\'{n}sk,  Institute of Theoretical Physics and Astrophysics and National Quantum Information Centre in Gda\'{n}sk, 81-824 Sopot, Poland \\
$^2$ Institute Of Advanced Study In Mathematics, Harbin Institute Of Technology, Harbin 150006, China\\
$^3$ Center for Theoretical Physics, Massachusetts Institute of Technology, Cambridge MA 02139, USA.}
%$^4$ Department of Computer Science and Department of Physics and Astronomy, University College London, WC1E 6BT London, U.K}

%\affiliation{
%$^1$ Institute of Theoretical Physics and Astrophysics, University of Gda\'nsk, 80-952 Gda\'nsk, Poland \\
%$^2$ School of Mathematics and Statistics, Wuhan University, China \\
%$^3$ National Quantum Information Centre of Gda\'nsk, 81-824 Sopot, Poland\\
%$^4$ Center for Theoretical Physics, Massachusetts Institute of Technology, US\\
%}

%%%%%%%%%%%%%%%%%%%%%

\begin{abstract}
  Maximal violation of the CHSH-Bell inequality is usually said to be a feature of
  anticommuting observables.  In this work we show that even random observables exhibit
  near-maximal violations of the CHSH-Bell inequality.  To do this, we use the tools of
  free probability theory to analyze the commutators of large random matrices.  Along the
  way, we introduce the notion of ``free observables'' which can be thought of as
  infinite-dimensional operators that reproduce the statistics of random matrices as their
  dimension tends towards infinity.  We also study the fine-grained uncertainty of a
  sequence of free or random observables, and use this to construct a steering inequality
  with a large violation.
\end{abstract}

%%%%%%%%%%%%%%%%%%%%%

%\keywords{Bell inequality, Steering inequality, Nonlocal correlations, free probability}

\maketitle

%%%%%%%%%%%%%%%%%%%%%%%%%%%%%%%%%%
%%%%%%%%%%%%%%%%%%%%%%%%%%%%%%%%%%

\section{Introduction}

The notion of quantum mechanics violating local realism was first raised by the work of
A.~Einstein, B.~Podolsky and N.~Rosen \cite{EPR}.  This was put on a rigorous and general
footing by the revolutionary 1964 paper of J.~S.~Bell~\cite{Bell}, which derived an
inequality (now known as the Bell inequality) involving correlations of two
observables. Bell showed that there is a constraint on any possible correlations obtained
from local hidden variable models which can be violated by quantum measurements of
entangled states.  Later on, another Bell-type inequality which is more experimentally
feasible was derived by J.~F.~Clauser, M.~A.~Horne, A. Shimony and R. A. Holt
\cite{CHSH1969}. Since then, Bell inequalities have played a fundamental role in quantum
theory and have had  applications in quantum information science including
cryptography, distributed computing, randomness generation and many others (see
\cite{BCPSW2014} for a review).

In this paper, we mainly focus on the maximal violation of CHSH-Bell inequality
\cite{CHSH1969}.  It is well known that the {\it Tsirelson bound} $2\sqrt{2}$ for CHSH-Bell inequality was first obtained by Tsirelson \cite{Tsirelson1980}. And he also proved that the bound can be realized by using proper Pauli observables. Apart from the above qubit case, it is possible to find dichotomic
observables in high dimension \cite{GP1992,BMR1992}, as well as in the continuous-variable
(infinite dimension) case \cite{CPHZ2002}, to obtain the Tsirelson bound. Recently,
Y.~C.~Liang et.~al.~\cite{LHBR2010} have studied the possibility of violation of CHSH-Bell
inequality by random observables. For the bipartite qubits case, if two observers share a Bell
state showed that random measure settings lead to a violation with probability $\approx
0.283$. However, for two qubits, the probability of the maximal
violation is zero, and the probability of near-maximal violation is negligible.

%So it is natural to ask following question: {\it Is it possible to maximally violate the
% CHSH-Bell inequality by using random observables with high probability?} In this work,
% we will provide an affirmative answer to this question by using random observables in
% sufficient large dimensional Hilbert space.
%Thus in this asymptotic sense, maximally violating CHSH-Bell inequality is {\it generic} for random observables.

Contrary to the case of qubits, our results show that the probability of near-maximal violation is large in high dimension. here near-maximal violations are approximately achieved with high probability by random high-dimensional observables. Previous methods of showing maximal violation were based on specific algebraic relations,
namely, anti-commuting, and indeed there is a sense in which maximal violations imply
anti-commutation on some subspace~\cite{MYS2012}.
However, this random approach reveals that there is another type of
algebraic relations between observables which might lead to the Tsirelson bound of
CHSH-Bell inequality. We call the observables, which satisfy those relations, {\it free
  observables.}   This terminology is from a mathematical theory called free probability
\cite{VDN1992,NS2006}. As we explain below, those free
observables are freely independent in some quantum probability space, which is a quantum analogue of the classical probability space (see the section IV for the definition).  A crucial point is
that free observables can only exist in infinite dimension, and thus are experimentally
infeasible.  We also discuss finite-dimensional approximations (section IV.B) which are more
experimentally plausible and for which
the Tsirelson bound can be approximately obtained.

In another part of this work we study the fine-grained uncertainty relations of free or
random observables, which was introduced by J. Oppenheim and S. Wehner \cite{OW2010}. It
is more fundamental than the usual entropic uncertainty relations and it relates to the
degree of violation of Bell inequalities (non-local games) \cite{OW2010,RGMH2015}. For a
pair of free (random) observables, we can show that the degree of their uncertainty is
0. On the other hand, it is interesting that for a sequence of free (random) observables
$A_1,\ldots,A_n$ with $n>4,$ the fine-grained uncertainty is upper bounded by $\frac{1}{2}
+ \frac{1}{\sqrt{n}},$ which is the same as the one given by the anti-commuting
observables. Therefore as a byproduct of above results, by using free (random) observables
we can obtain one type of steering inequality with large violation that recently was
studied in \cite{MRYHH2015}.

%%%%%%%%%%%%%%%%%%%%%%%%%%%%%%%%%%%

\section{Preliminaries}

%%%%%%%%%%%%%%%%%%%%%%%%%%%%%%%%%%%

First, we introduce terminology.  For a bipartite dichotomic Bell scenario,  there are two
space-like separated observers, say, Alice and Bob. Each of them is described by a
$N$-dimensional Hilbert space $H_N,$ and Alice (resp.~Bob) chooses one of $n$ dichotomic
(i.e.~two-outcome) observables $A_i$ (resp. $B_j$) that will take results $\alpha_i$
(resp. $\beta_j$) from set $\{1, -1\}.$  Thus the observables are self-adjoint unitaries.

Next, recall the famous CHSH-Bell inequality \cite{CHSH1969}.  If
$\alpha_1,\alpha_2,\beta_1,\beta_2$ are classically correlated random variables then
\begin{equation}
\left| \langle \alpha_1 \beta_1 \rangle + \langle \alpha_1 \beta_2 \rangle + \langle \alpha_2 \beta_1 \rangle - \langle \alpha_2 \beta_2 \rangle \right| \leq 2,
\end{equation}
so we say that 2 is the largest {\it classical value}  obtained by any local hidden
variable model. In \cite{Tsirelson1980}, Tsirelson first proved that if the correlations
are obtained by quantum theory then the {\it quantum value} of the CHSH-Bell inequality is
$2\sqrt{2}$ (i.e., the Tsirelson bound). To see this, consider the following CHSH-Bell operator
\begin{equation}
B_{\text{CHSH}} = A_1 \otimes B_1 + A_1 \otimes B_2 + A_2 \otimes B_1 - A_2 \otimes B_2,
\end{equation}
where $A_i, B_j, i,j =1,2$ are dichotomic observables. By choosing proper observables,
e.g. $A_1 = \sigma_x, A_2 = \sigma_z, B_1 = (\sigma_x + \sigma_z) /\sqrt{2}, B_2 =
(\sigma_x - \sigma_z)/\sqrt{2},$ the norm (largest singular value) of the CHSH-Bell
operator is $2\sqrt{2}.$
If $\mathcal{B} = B_{\text{CHSH}}^2,$ then
\begin{equation}\label{eq:B^2}
\mathcal{B} = 4 \un - [A_1, A_2] \otimes [B_1, B_2].
\end{equation}
If both parties choose compatible (commutative) observables, then $\mathcal{B} = 4 \un.$
Hence incompatible (non-commutative) observables are necessary for the violation of
CHSH-Bell inequality \cite{BMR1992}.  The Tsirelson bound is also determined by
the eigenvalues of the commutators $[A_1, A_2]$ and $[B_1, B_2].$ More precisely, suppose
the local dimension for each party is $N,$ and the eigenvalues of $[A_1, A_2]$
(resp. $[B_1, B_2]$) are $s_1,\ldots, s_N$ (resp. $t_1, \ldots, t_N$). Then we have
\cite{BMR1992}:
\begin{equation}
\|\mathcal{B} \| = \max_{i,j} \{ 4 - s_i t_j \}.
\end{equation}
It is clear that if there exist eigenstates such that the eigenvalues of $[A_1, A_2]$
(resp. $[B_1, B_2]$) are $\pm2,$ then $\|B_{\text{CHSH}}\| = 2\sqrt{2}.$  In particular,
anti-commuting dichotomic local observables, such as $\sigma_x$ and $\sigma_z$, will
saturate the Tsirelson bound.

%%%%%%%%%%%%%%%%%%%%%%%%%%%%%%%

\section{A random approach to the Tsirelson bound}\label{sec-random}

%%%%%%%%%%%%%%%%%%%%%%%%%%%%%%%

Suppose $D$ is a $N\times N$ deterministic diagonal matrix, where the diagonal terms of
$D$ are either $1$ or $-1$ and $\Tr (D) =0$ where $Tr$ is the usual trace for matrices. It is easy to see that $D^2 = \un.$ Suppose
unitaries $U_i, i=1,\ldots, n$ are independent Haar-random matrices in the group of
unitary matrices $U(N).$
Define the following random dichotomic observables:
\begin{equation}
A_i = U_i D U_i^\dagger, \; i=1, \ldots, n.
\end{equation}
We would like to establish results that hold with ``high
probability'' over some natural distribution.
 Recall that we call a sequence of random variables $\{X_N\}_N$ convergent
to $X$ almost surely in probability space $(\Omega, P),$ if $P \left( \lim_{N\to \infty}
  X_N = X \right) =1.$ With these notions, we claim that the Tsirelson bound of CHSH-Bell
inequality can be obtained in high probability by using random dichotomic observables in
sufficient large dimension. More precisely, we have following theorem:

\begin{theorem}\label{thm:Main}
Let $A_i = U_i D U_i^\dagger$ and $B_i = V_i D V_i^\dagger, i=1,2,$ where $U_i, V_i$ are
independent Haar-random unitaries in $U (N).$ Then we have
\begin{equation}
\lim_{N \to \infty} \|B_{\text{CHSH}}\| = 2\sqrt{2},  \;\; \text{almost surely}.
\end{equation}
\end{theorem}
 Above theorem could be understand as the following: with sufficient large dimension, the random dichotomic observables may saturate the Tsirelson bound of the CHSH-Bell inequality. We note here that in this approximate scenario, the shared state for Alice and Bob should not be fixed, otherwise it may not obtain any violation at all.
To prove this theorem, we first need following lemma from \cite{NS2006}:
\begin{lem}\label{lem:c-norm}[\cite{NS2006}]
Let $\mathcal{M}_N$ be the set of $N \times N$ matrices. Then for every $ A \in \mathcal{M}_N,$
\begin{equation}
\|A\| = \lim_{k\rightarrow \infty} \left( \tr_N \left((A^{\dagger} A)^k \right)\right)^{\frac{1}{2k}},
\end{equation}
where $\tr_N = \Tr/N.$
\end{lem}

Now denote $A= [A_1, A_2]$ and $B = [B_1, B_2].$  For any $k \in \mathbb{N}_0,$ by we can
use the binomial formula and equation \eqref{eq:B^2} to obtain
\begin{equation}
\begin{split}
\tr_{N^2} (\mathcal{B}^k) & = \tr_{N^2}(4\un - A \otimes B)^k \\
& =\sum_{j=0}^{k}\binom{k}{j} 4^{k-j} (-1)^{j} \tr_N (A^j) \cdot  \tr_N (B^j).
\end{split}
\end{equation}

Let us consider the term $\tr (A^j).$ Since $A_1A_2$ and $A_2A_1$ commute, again by binomial formula, we have
\begin{equation}
\tr_N (A^j) = \sum_{l=0}^j \binom{j}{l} (-1)^{j-l} \tr_N ((A_1A_2)^{|2l-j|}).
\end{equation}

Now we need the second key lemma (see Appendix B for the details of proof).
\begin{lem}\label{lem:key}
Let $A_i = U_i D U_i^\dagger,$ where $U_i, i=1,\ldots, n \in U(N)$ are independent Haar
random unitaries.  Consider a sequence
$i(1), \ldots, i(k) \in [n]$ satisfying $i(1) \neq i(2) \neq i(3) \neq \ldots i(k-1) \neq i(k).$
Then
\begin{equation}\label{eq:asym-free}
\lim_{N \to \infty} \tr_N (A_{i(1)}A_{i(2)}\cdots A_{i(k)}) = 0, \;\; \text{almost surely}.
\end{equation}
\end{lem}

 This lemma is mostly due to the work of B. Collins~\cite{Collins2002,CS2004}, where he
 and other co-authors developed a method to calculate the moments of polynomial random
 variables on unitary groups.  This method is called the Weingarten calculus and is in
 turn based on \cite{Weingarten1978}.  As we will see in the next section, this lemma  can
 be thought of as establishing the ``asymptotic freeness''
 of these random matrices.  Thus by Lemma \ref{lem:key},  we have  (almost surely)
\begin{equation}
\begin{split}
 \lim_{N \to \infty} \tr_N (A^j)& =  \sum_{l=0}^j \binom{j}{l} (-1)^{j-l} \lim_{N \to \infty} \tr_N ((A_1A_2)^{|2l-j|})\\
 & = \left \{ \begin{split}
 & (-1)^{j/2} \binom{j}{j/2} , \;\; \text{j is even},\\
 & 0, \;\; \text{otherwise}.
 \end{split} \right.
\end{split}
\end{equation}
A similar estimate is also valid for the term $\tr_N (B^j).$ Therefore
\begin{equation}\label{eq:fact}
\lim_{N \to \infty} \tr_{N^2} (\mathcal{B}^k)  = \sum_{j=0, \; \text{j is even}}^{k}\binom{k}{j} 4^{k-j} \binom{j}{j/2}^2 : = Q_k, \;\;  \text{almost surely}.
\end{equation}
By Stirling's formula, we have $\lim_{k \to \infty} (Q_{2k})^{1/2k} =8.$
In other words, for any $\epsilon>0,$ we can choose $k \in \mathbb{N},$ such that
$(Q_{2k})^{1/2k} > 8- \epsilon.$
Since $( \tr_{N^2} \mathcal{B}^{k})^{1/k} \leq \|\mathcal{B}\|$ for all $k \geq 1,$ then we have
\begin{equation}
\liminf_{N \to \infty} \|\mathcal{B}\| \geq (Q_{2k})^{1/2k} > 8- \epsilon, \;\; \text{almost surely.}
\end{equation}
On the other hand, due to Tsirelson's inequality~\cite{Tsirelson1980} we have $\|\mathcal{B}\| \leq 8$. Thus we complete our proof of Theorem \ref{thm:Main}.

%%%%%%%%%%%%%%%%%%%%%%%%%%%%%%%%%

\section{A free approach to the Tsirelson bound}

%%%%%%%%%%%%%%%%%%%%%%%%%%%%%%%%%

The random dichotomic observables do not satisfy the anti-commuting relations. In fact,  random dichotomic observables are ``asymptotically'' freely independent, which
was first established by Voiculescu \cite{V1991} in the case of the Gaussian unitary ensemble
(GUE). That result builds a gorgeous bridge across two distinct mathematical
branches--random matrix theory and free probability. In free probability theory, we will
treat observables $A_i, B_j$ as elements of a $C^*$-algebra $\mathcal{A},$ equipped with
an unital (faithful) state $\phi$, where ``state'' means a linear map from $\mathcal{A}$
to $\mathbb{R}$, unital means $\phi(\un) =1$ and faithful means
$\phi(AA^*)=0 \Rightarrow A =0.$ The pair $(\mathcal{A}, \phi)$ is called a
$C^*$-probability space, which is a quantum analogue of a classical probability space and we can call it a "quantum" probability space. For
example, $(\mathcal{M}_N, \tr_N)$ is a $C^*$-probability space, where $\mathcal{M}_N$ is the
set of $N \times N$ matrices. We refer to \cite{NS2006} for more details of quantum probability.

Lemma \ref{lem:key} inspires us to consider the following adaptation of definition of freeness to the case of dichotomic observables.
\begin{definition}\label{def:freeness}
For given $C^*$-probability space $(\mathcal{A}, \phi),$ dichotomic observables $A_i, i\in I$ are called freely independent, if
\begin{equation}
\phi (A_{i(1)}A_{i(2)}\cdots A_{i(k)}) =0
\end{equation}
whenever we have following:
\begin{enumerate}[{\rm (i)}]
\item $k$ is positive; $i(1), i(2), \ldots, i(k) \in I$;
\item $\phi(A_{i(k)}) =0$ for all $k$;
\item $i(1) \neq i(2), i(2) \neq i(3), \ldots, i(k-1) \neq i(k).$
\end{enumerate}
\end{definition}

For the special case $I= \{1,2\},$ the above conditions are equivalent to
\begin{equation}\label{eq:two-free}
\phi (A_1) = \phi(A_2) = \phi (A_1A_2) = \phi (A_2A_1) = \phi(A_1A_2A_1) = \phi(A_2A_1A_2) = \cdots = 0.
\end{equation}

However, finite-dimensional
observables cannot be freely independent.
In other words, for fixed $N,$ the $C^*$-probability space $(M_N, \tr_N)$ is too small to
talk about freeness, and Definition \ref{def:freeness} refers to an empty set.
Fortunately if we consider the observables in infinite dimensional Hilbert space, it is
possible for them to be freely independent in some $C^*$-probability space $(\mathcal{A},
\phi).$ Furthermore, the derivations in Section~\ref{sec-random} do not depend on
the dimension.  In order to use an infinite dimensional
$C^*$-probability space $(\mathcal{A},\phi)$ instead of $(M_N, \tr),$ we need only update
Lemma \ref{lem:c-norm} with an appropriate formula, which is achieved by \eqref{eq:norm}
below.  We conclude as follows.

\begin{theorem}\label{thm:main}
For the CHSH-Bell inequality, the Tsirelson bound can be obtained by using observables which are freely independent in their respective local system. More precisely, if $A_1, A_2$ and $B_1, B_2$ are freely independent in some $C^*$-probability space $(\mathcal{A}, \phi),$ then we have
$ \|B_{\text{CHSH}}\| = 2\sqrt{2}.$
\end{theorem}
This result is rather abstract, but in the next subsection, we will provide a concrete
example which satisfies the conditions in this theorem.

%%%%%%%%%%%%%%%%%%%%%%%%%%%%%%%%%%

\subsection{A concrete example in infinite dimension}\label{sub-example}

%%%%%%%%%%%%%%%%%%%%%%%%%%%%%%%%%%

For infinite-dimensional $C^*$-probability space, Definition \ref{def:freeness} is
meaningful. Now consider a group $G= \ast_n \mathbb{Z}_2$ and its associated Hilbert space
$\ell_2(G).$ This notation refers to the $n$-fold free product of $\mathbb{Z}_2$ with
itself; i.e.~the infinite group  $G$ with the  the following elements: $g_{i_1}, g_{i_1}g_{i_2}, \ldots, g_{i_1}g_{i_2}\cdots g_{i_n}, i_1, \ldots, i_n =1, \ldots, n,$ where $g_1, \ldots, g_n$ are the generators of the group $G$  whose only
relations are $g_i^2=1$.
The set $\{|g\rangle: g\in G\}$ forms an orthonormal basis of $\ell_2(G),$
thus the dimension of $\ell_2(G)$ is infinite. Let $\lambda : G \rightarrow
B(\ell^2(G))$ be the left regular group representation, which is defined as:
\begin{equation}
\lambda(g) |h \rangle = |gh\rangle, \; \forall h \in G.
\end{equation}
The reduced $C^*$-algebra $C^*_{\text{red}}(G)$ is defined as the norm closure of the linear span
$\{\lambda(g), g\in G\},$ where the norm is the operator norm of $B(\ell_2(G)).$ There is
a faithful trace state $\phi$ on $C^*_{\text{red}}(G)$ defined as
\begin{equation}\label{eq:state}
\phi\Big(\sum_g \alpha_g \lambda(g)\Big) := \alpha_e.
\end{equation}
Obviously $\phi(\un) =1.$ Hence $(C^*_{\text{red}}(G),\phi)$ is a $C^*$-probability space. If
$g_i$ is the generator of  the $i$-th copy of $\ast_n \mathbb{Z}_2, i=1,2,\ldots, n$ then
\begin{equation}
A_i = \lambda(g_i), \;\; i=1, \ldots, n,
\end{equation}
It is easy to check $A_i, i=1,\ldots, n$ are self-adjoint unitaries and freely independent
in $(C^*_{\text{red}}(G), \phi)$.   We will choose  the local Hilbert spaces of Alice and Bob  to
be $\ell_2(G),$ where $n=2.$ By using those free observables, we can obtain the quantum
value $2\sqrt{2}$ for CHSH-Bell inequality.
Note conjugating by a unitary preserves
freeness of observables, i.e, if $A_1, A_2$ are freely independent, then $U A_1 U^\dagger,
U A_2 U^\dagger$ are still freely independent for any unitary $U$. Since the norm of Bell
operator does not change under the local unitary operation, we can simply assume $A_1=B_1
= \lambda(g_1), A_2 = B_2 = \lambda(g_2).$

%%%%%%%%%%%%%%%%%%%%%%%%%%%%%%%%%
%%%%%%%%%%%%%%%%%%%%%%%%%%%%%%%%%

\subsection{Truncated free observables in finite dimension}\label{sub-free-for-poor}

In order to see how the freeness behaves in a simple and direct way, we will truncate the
free observables given by last subsection to finite dimension. Denote the elements in
$\ell_2(\ast_2 \mathbb{Z}_2)$ as follows:
\begin{equation}
 \begin{array}{ccccccccc} \cdots & \left|g_2g_1g_2\right\rangle & \left|g_2g_1\right\rangle & \left|g_2\right\rangle & \left|e\right\rangle & \left|g_1\right\rangle & \left|g_1g_2\right\rangle & \left|g_1g_2g_1\right\rangle & \cdots \\ \updownarrow & \updownarrow & \updownarrow & \updownarrow & \updownarrow & \updownarrow & \updownarrow & \updownarrow & \updownarrow \\ \cdots & \left|-3\right\rangle & \left|-2\right\rangle & \left|-1\right\rangle & \left|0\right\rangle & \left|1\right\rangle & \left|2\right\rangle & \left|3\right\rangle & \cdots \end{array}.
\end{equation}
With the above notation, we have
\begin{equation}
 \left \{ \begin{split}
 & \lambda(g_1)|i\rangle = |j\rangle, \; i+j=1,\\
 & \lambda(g_2)|i\rangle = |j\rangle, \; i+j=-1.
 \end{split} \right.
\end{equation}
where $i, j = \cdots, -1, 0, 1, \cdots$.

Now define $A_1^{(N)}$ and $A_2^{(N)}$ to be the truncation of the free observables to dimension $N = 2l+1$ (i.e, we truncated the operators $\lambda(g_1), \lambda(g_2)$ into the operators acting on $N$ dimension Hilbert space.). Then we have (see Figure 1):
\begin{subequations}\label{eq:cut1}\begin{align}
  A_1^{(N)}|i\rangle &= |1-i\rangle, &  i=-l+1, \ldots, l\\
 A_1^{(N)}|-l\rangle &= |-l\rangle, &
\end{align}
\end{subequations}
and
\begin{subequations}\label{eq:cut2}\begin{align}
A_2^{(N)}|i\rangle &= |-1-i\rangle, & i=-l, \ldots, l-1\\
  A_2^{(N)}|l\rangle &= |l\rangle.
\end{align}
\end{subequations}
where $|i \rangle, i= -l, \ldots, l$ denotes the basis of the $N$ dimensional Hilbert space.

\begin{figure}[htbp]
\centering
\includegraphics[height=4cm]{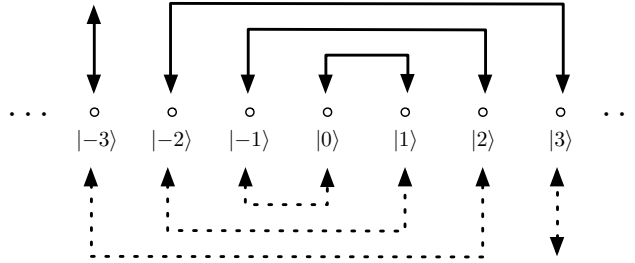}
\caption{The solid line stands for $A_1^{(N)}$ and the dashed lines stands for $A_2^{(N)}$, where $N=7$. For example, the operator $A_1^{(N)}$ maps the vector $|0\rangle$ to $|1\rangle$ and vice versa. We only need to be careful for the vector $|-3\rangle$, where $A_1^N$ maps it to itself.}
\end{figure}

It is clear that $A_1^{(N)}$ and $A_2^{(N)}$ are self-adjoint unitaries. Thus they can be
treated as a pair of dichotomic observables in an $N$-dimensional Hilbert space. Denote
$S= A_2^{(N)} \circ A_1^{(N)},$ so that
\begin{subequations}\label{eq:cut3}\begin{align}
  S |j\rangle &= |j-2\rangle, & j=-l+2,\ldots, l\\
  S |-l+1\rangle& = |l\rangle,\\
 S |-l\rangle& = |l-1\rangle.
\end{align}\end{subequations}
By the following diagram it is easy to see that $S$ is a cycle in the permutation group $S_{N}.$
\begin{equation}
\begin{array}{ccccccccccc}
|l\ra & \rar & |l-2\ra & \rar & |l-4\ra & \rar & \cdots & \rar & |-l+2\ra & \rar & |-l\ra \\
\nwarrow &&&&&&&&&& \swarrow \\
& |-l+1\ra & \lar & |-l+3\ra & \lar & \cdots & \lar & |l-3\ra & \lar & |l-1\ra. &
\end{array}
\end{equation}

Now for the CHSH-Bell operator $B_{\text{CHSH}},$ by using those truncated free observables, we can show that the quantum value tends to $2\sqrt{2}$ as $N \rightarrow \infty.$ Then due to the fact that the eigenvalues of $S$ are $\lambda_j = \exp^{2 \pi i j/N}, j=0, \ldots, N-1,$ we have
\begin{equation}
\begin{split}
\|B_{\text{CHSH}}^2\| & = \left\| 4 \un - [A_1^{(N)}, A_2^{(N)}] \otimes [A_1^{(N)}, A_2^{(N)}]\right\|\\
& = \left\| 4 \un - (S^\dagger - S) \otimes (S^\dagger -S) \right\| \\
& = \max_j \{ 4 + 4 (\Im(\lambda_j))^2 \} = \max_j \left\{ 4+ 4 \sin^2{\frac{2\pi j}{N}} \right\}\\
& \thickapprox 4 + 4 \left(1- O\left(\frac{1}{N^2}\right)\right) = 8 - O\left( \frac{1}{N^2} \right).
\end{split}
\end{equation}

%\begin{remark}
%Here we have only considered the case for truncation in odd dimension. For the even dimensional case, we just assume $A^{(2N+2)} = A^{(2N+1)} \oplus 0.$
%\end{remark}

Here for simplicity, we have assumed that Alice and Bob take same measurements. Therefore, we have following proposition:
\begin{proposition}\label{prop:cut}
By using truncated free observables $A_1^{(N)}, A_2^{(N)}, N= 2l+1,$ we can asymptotically obtain the Tsirelson bound for CHSH-Bell inequality, i.e,
$\|B_{\text{CHSH}}\| = 2\sqrt{2} - O(1/N^2).$
\end{proposition}

This result suggests the speed of the convergence mentioned in Theorem \ref{thm:Main}, namely, the Tsirelson bound will be saturated with the speed of $O(1/N)$ by using the random observables. However, the rigorous proof would need very careful and subtle analysis of Weingarten calculus.

%%%%%%%%%%%%%%%%%%%%%%%%%%%%%%%%%%%%%%
%%%%%%%%%%%%%%%%%%%%%%%%%%%%%%%%%%%%%%

\section{Fine-grained uncertainty relations for random (free) observables}

The uncertainty principle and non-locality are two fundamental and intrinsic concepts of
quantum theory which were quantitatively linked by J. Oppenheim and S. Wehner's work
\cite{OW2010}.  There they introduced a notion called ``fine-grained uncertainty relations''
to quantify the ``amount of uncertainty'' in a particular physical theory.  Suppose we have $n$ dichotomic observables $A_i, i=1,\ldots,n,$
corresponding to measurement settings  $P_i^a = \frac{\un + (-1)^a A_i}{2}, i=1,\ldots,n;
a=0,1.$ The uncertainty of measurement settings $P_i^0, i=1, \ldots, n$ is defined as:
\begin{equation}
\xi_{\vec{0}} = \sup_{\rho} \left \{ \frac{1}{n} \sum_{i=1}^n \Tr (P_i^0 \rho)\right\} = \frac{1}{2} + \frac{1}{2n} \sup_{\rho} \Tr \left( \sum_{i=1}^n A_i \rho \right).
\end{equation}
Similarly, the uncertainty of $P_i^1, i=1,\ldots,n$ is
\begin{equation}
\xi_{\vec{1}} = \sup_{\rho} \left \{\frac{1}{n} \sum_{i=1}^n \Tr (P_i^1 \rho)\right\} = \frac{1}{2} - \frac{1}{2n} \sup_{\rho} \Tr \left( \sum_{i=1}^n A_i \rho \right).
\end{equation}
Notice that
\begin{equation}
\sup_{\rho} \left| \Tr \left( \sum_{i=1}^n A_i \rho \right) \right| = \left\| \sum_{i=1}^n A_i \right\|.
\end{equation}
Hence $\xi_{\vec{0}}= \xi_{\vec{1}} = \frac{1}{2} + \frac{1}{2n} \left\| \sum_{i=1}^n A_i \right\|.$  The state $\rho$ which can obtain $\xi_{\vec{x}}$ is called the maximally certain state for those measurement settings. If we assume $A_i$ are freely independent observables, then we have following proposition (see Appendix C and D for the proof):
\begin{proposition}\label{prop:uncertainty-free}
The fine-grained uncertainty for free observables $A_i, i=1, \ldots, n, n>4$ is
\begin{equation}
\xi_{\vec{0}}= \xi_{\vec{1}}  \leq \frac{1}{2} + \frac{1}{\sqrt{n}}< 1.
\end{equation}
The same results approximately hold for random observables $A_i = U_i D U_i^\dagger,
i=1,\ldots, n, n>4$ with high probability.
\end{proposition}

For the special case $n=2,$ we have $\|A_1 + A_2 \| =2.$
Thus for $n=2,$ $\xi_{\vec{0}}= \xi_{\vec{1}} =1$ (see Appendix D for random observables and Appendix E for free obervables). Interestingly, for truncated free observables we have
\begin{equation}
\begin{split}
\|A_1^{(N)} + A_2^{(N)}\|^2 & = 2 \un + S + S^\dagger = \max_j \{2 + 2 \Re(\lambda_j)\} \\
& = \max_j \left\{2 + 2 \cos{\frac{2\pi j}{N}} \right\} = 4.
\end{split}
\end{equation}
Thus for the truncated free observables, we always have $\xi_{\vec{0}} =\xi_{\vec{1}} =1$ regardless of what dimension we truncate to.

In a recent work, some of us show that there is a tight relationship between fine-grained
uncertainty and violation of one specific steering inequality, called the linear steering
inequality, which was first used in \cite{SJWP2010} to verify steering by experiment. It
has following form:
\begin{equation}\label{eq:steering-ineq}
S_n = \sum_{i=1}^n \langle \alpha_i A_i \rangle \leq  C_n,
\end{equation}
where $C_n$ is called the {\it local hidden state} bound of $S_n.$ This bound can be calculated easily as follows \cite{SJWP2010}:
\begin{equation}\label{eq:LHS-bound}
C_n = \sup_{\alpha_i= \pm 1} \left\| \sum_{i=1}^n \alpha_i A_i  \right\|.
\end{equation}
If the observables $A_i$ are chosen to be operators of a Clifford algebra, which are
anti-commutative, a large (unbounded) violation can be obtained \cite{MRYHH2015}. Because
the degree of the fine-grained uncertainty of free or random observables is the same
order as that of anti-commuting observables,  we find:
\begin{corollary}
If $A_i, i=1, \ldots, n$ are chosen to be free observables, then the local hidden state bound of steering inequality $S_n = \sum_{i=1}^n \langle \alpha_i A_i \rangle \leq  C_n,$ is upper bounded by $2\sqrt{n}.$ The similar result holds for random observables with high probability.
\end{corollary}

Here we note that for the free case, we should also care about the quantum values of
steering inequalities. Due to M. Navascu\'{e}s and D. P\'{e}rez-Garc\'{i}a's work, there
are two different ways to define them~\cite{NG2012}. One is in a commuting way that means
the system described by a total Hilbert space, and the other one is the total system
described in a tensor form.  As a matter of fact, they also used the free observables
$\lambda(g_i), i=1, \ldots, n$ to define the linear steering inequality. They showed the
quantum value defined in the commuting sense is $n,$ while in the tensor scenario is upper
bounded by $2\sqrt{n-1}.$ So by their work, we can easily see that the local hidden state
bound is upper bounded by $2\sqrt{n-1}$ for free observables. Their bound is even sharper
than ours. However, we have provided another proof which is more focussed  on the freeness
property and is applicable to random observables.

%%%%%%%%%%%%%%%%%%%%%%%%%%%%%%%%%%%%%%
%%%%%%%%%%%%%%%%%%%%%%%%%%%%%%%%%%%%%%

\section{Conclusions}

In this paper, we show that random dichotomic observables generically achieve near-maximal violation
of the CHSH-Bell inequality, approaching the Tsirelson bound in the limit of large
dimension.
  This is despite the fact that these observables are not  anti-commuting.
Instead, due to Voiculescu's theory, they are asymptotically freely independent. It means
when the dimension increases, their behaviors tends to the ones of free observables in
some quantum probability space.  However, the quantum state that is optimal for the random observables is random as well, as it in general will depend on the observables. For a fixed state, random observables might not lead to any violation. Another main result of this paper is that we have considered the fine-grained uncertainty of a sequence of free or random observables. The
degree of their uncertainty is as the same order as the one which is given by the
anti-commuting observables. As a byproduct of this result, we can construct a linear
steering inequality with large violation by using free or random observables. For further
applications, free observables may be used for studying the quantum value of other type of
Bell inequalities. Thus a natural question arises: {\it Do free observables always
  maximally violate any Bell inequalities?} Unfortunately, a quick answer is that we can
consider the linear Bell operator $ \sum_{i=1}^n  A_i\otimes  A_i.$ It is trivial since
its quantum and classical value are both $n,$ while the quantum value given by free
observables is upper bounded by $2\sqrt{n}$. However, it seems promising when considering
other specific Bell inequalities. Since the free observables and their truncated ones are
deterministic (constructive), another possible application is that this may be a new
source of constructive examples of Bell inequality violations where previously only random
ones were known.

\vspace{3mm}

{\it Acknowledgments}---We would like to thank Pawe{\l} Horodecki, Marek Bozejko, Mikael de la Salle, Yanqi Qiu and Junghee Ryu for valuable discussions. We also would like to thank the anonymous referee for her/his useful comments.
This work is supported by ERC AdG QOLAPS, EU grant RAQUEL and the Foundation for
Polish Science TEAM project cofinanced by the EU European Regional Development Fund. Z. Yin was
partly supported by NSFC under Grant No.11301401. A. Rutkowski was supported by a postdoc internship decision number DEC\textendash{} 2012/04/S/ST2/00002 and grant No. 2014/14/M/ST2/00818 from the Polish National Science Center. M. Marciniak was supported by EU project BRISQ2. Harrow was funded by NSF grants CCF-1111382 and CCF-1452616,  ARO contract
W911NF-12-1-0486 and a Leverhulme Trust Visiting Professorship VP2-2013-041.

%%%%%%%%%%%%%%%%%%%%%%%%%%%%%%%%%%%%%%

\bibliographystyle{plain}

%%%%%%%%%%%%%%%%%%%%%%%%%%%%
%%%%%%%%%%%%%%%%%%%%%%%%%%%%

\section*{Appendix}

\subsection{$C^*$-probability space and freely independent}

\begin{definition}
A $\ast$-probability space $(\mathcal{A}, \phi)$ consists of an unital $\ast$-algebra $\mathcal{A}$ over $\mathbb{C}$ and an unital linear positive functional
\begin{equation}
\phi: \mathcal{A} \rightarrow  \mathbb{C} ; \;\;\; \phi(1_{\mathcal{A}}) = 1.
\end{equation}
The elements $a \in \mathcal{A}$ are called non-commutative random variables
in $(\mathcal{A}, \phi).$ A $C^*$-probability space is a $\ast$-probability space
$(\mathcal{A}, \phi)$ where $\mathcal{A}$ is an unital $C^*$-algebra.
\end{definition}
If additionally we assume $\phi$ is faithful, we have for any $a \in \mathcal{A},$
\begin{equation}\label{eq:norm}
\|a\| = \lim_{k\rightarrow \infty} \left( \phi \left((a^* a)^k \right)\right)^{\frac{1}{2k}}.
\end{equation}

\begin{definition}\cite{NS2006,BM2013}
For given $C^*$-probability space $(\mathcal{A}, \varphi),$ let $\mathcal{A}_1, \ldots, \mathcal{A}_n$ be $\ast$-subalgebras of $\mathcal{A}$. They are said to be free if for all $a_i \in \mathcal{A}_{j(i)}, i = 1,\ldots n,\; j(i) \in \{1, \ldots, n\}$ such that $\phi(a_i) = 0$, one has
\begin{equation}
\phi (a_1a_2\cdots a_n) =0
\end{equation}
whenever $j(1) \neq j(2), j(2) \neq j(3), \ldots, j(n-1) \neq j(n).$ A sequence of random variables are said to be free if
the unital subalgebras they generate are free.
\end{definition}

\subsection{Proofs for Lemma \ref{lem:key}}

Lemma \ref{lem:key} is a direct corollary of the work of B. Collins \cite{Collins2002}. A random variable $u \in (\mathcal{A}, \phi)$ is called a Haar unitary when it is unitary and
\begin{equation}\label{eq:Haar}
 \phi (u^j) = \left \{ \begin{split}
 & 1 , \;\; j=0,\\
 & 0, \;\; \text{otherwise}.
 \end{split} \right.
\end{equation}
Since we have
\begin{equation}
\lim_{N \to \infty} \tr (D^j)= \left \{ \begin{split}
 & 1 , \;\; \text{j is even},\\
 & 0, \;\; \text{j is odd}.
 \end{split} \right.
\end{equation}
Then there will exist a $C^*$-probability space $(\mathcal{A}, \phi)$ and a random variable $d \in \mathcal{A},$ such that
\begin{equation}\label{eq:D}
\lim_{N \to \infty} \tr (D^j) = \phi (d^j), \;\; \text{for all}\; j\geq 0.
\end{equation}
Let $u_1, \ldots, u_n$ be a sequence of Haar unitaries in $(\mathcal{A}, \phi)$ which are freely independent together with $d$. We will give a concrete example of $u_1, \ldots, u_n, d$ in the end of this subsection. Let $E (\cdot)= \int \cdot \; d\mu,$ where $d\mu$ is the Haar measure on $U(N),$ then by the main theorem of \cite[Theorem 3.1]{Collins2002}, we have following:
\begin{equation}
\begin{split}
\lim_{N \to \infty} E \; \tr (A_{i(1)}A_{i(2)} \cdots A_{i(k)}) &= \phi (u_{i(1)}d u_{i(1)}^* \cdots u_{i(k)}d u_{i(k)}^*)\\
& = 0,
\end{split}
\end{equation}
where the second equation comes from the freeness of $d, u_1, \ldots, u_n.$ Moreover, by theorem of \cite[Theorem 3.5]{Collins2002}
\begin{equation}
P \left( \left| \tr (A_{i(1)}A_{i(2)} \cdots A_{i(k)}) \right| \geq \epsilon \right) = O(N^{-2}).
\end{equation}
Then by the Borel-Cantelli Lemma, for any $\epsilon>0,$
\begin{equation}
\limsup_{N \to \infty} \left| \tr (A_{i(1)}A_{i(2)} \cdots A_{i(k)}) \right| \leq \epsilon, \;\; \text{almost surely}.
\end{equation}
Hence
\begin{equation}
\lim_{N \to \infty}  \tr (A_{i(1)}A_{i(2)} \cdots A_{i(k)}) =0, \;\; \text{almost surely}.
\end{equation}

{\bf A concrete example of $u_1, \ldots, u_n$ and $d$.}

Let $G= \ast_{2n+1} \mathbb{Z}_2$ and $g_i, i=1, \ldots, 2n+1$ be the generator of the
$i$-th copy. Let $u_i= \lambda(g_{2i-1} g_{2i}), i=1,\ldots,n$ and $d=\lambda(g_{2n+1}).$
Then the $C^*$-probability we consider is $(C^*_{\text{red}}(G), \phi)$ which was defined
in Subsection \ref{sub-example}. It is easy to check that equations \eqref{eq:Haar} and \eqref{eq:D} hold. Thus $u_i, i=1, \ldots, n$ are Haar unitaries in $(C^*_{\text{red}}(G), \phi).$ Moreover $u_1, \ldots, u_n, d$ are freely independent in $(C^*_{\text{red}}(G), \phi).$

%%%%%%%%%%%%%%%%%%%%%%%%%%%%%%%%%

\subsection{Proof of Proposition \ref{prop:uncertainty-free}}

Suppose the dichotomic observables $A_i, i=1,\ldots, n$ are freely independent in some $C^*$-probability space $(\mathcal{A}, \phi)$. Then by equation \eqref{eq:norm},
\begin{equation}\label{eq:free-LHS-bound}
\begin{split}
\left\| \sum_{i=1}^n A_i \right\| &= \lim_{k\rightarrow \infty} \left(\phi \left( \sum_{i=1}^n  A_i \right)^{2k} \right)^{\frac{1}{2k}}\\
& =  \lim_{k\rightarrow \infty} \left(\phi \left( \sum_{i(1), \ldots, i(2k)=1}^n  A_{i(1)} \cdots A_{i(2k)} \right)\right)^{\frac{1}{2k}}.
\end{split}
\end{equation}

To estimate the above equation we need the following definitions and facts from
combinatorics~\cite{NS2006}. For a given set $\{1, \ldots, 2k\},$ there is a partition $\pi
= \{V_1, \ldots , V_s\}$ of this set. $\pi$ is determined as follows: Two numbers $p$ and
$q$ belong to the same block $V_k$ of $\pi$ if and only if $i(p) = i(q).$ There is a
particular partition called pair partition, in which every block only contains two
elements. A pair partition of $\{1, \ldots , 2k\}$ is called non-crossing if there does
not exist $1\leq  p_1 < q_1 < p_2 < q_2 \leq 2k$ such that $p_1$ is paired with $p_2$ and
$q_1$ is paired with $q_2.$ The number of non-crossing pair partitions of the set $\{1,
\ldots , 2k\}$ is given by the Catalan number $C_k= \frac{1}{k+1}\binom{2k}{k}$.

Now for the indices $i(1), \ldots, i(2k),$ if there exist a pair of adjacent indices which they belong to a same block, e.g. $i(s-1) = i(s),$ then we will shrink the indices $i(1), \ldots, i(2k)$ to $i(1), \ldots, i(s-2), \emptyset, i(s+1), \ldots, i(2k),$ since obviously $A_{i(s-1)}  A_{i(s)} = \un.$ According to this rule, we can shrink $\pi$ to a new partition $\tilde{\pi}$ on $\{1, \ldots, 2t\},$ where $t \leq k.$ Hence we can divide $\pi$ into two groups:

{\it Case 1.} $\tilde{\pi}= \emptyset$.

{\it Case $2.$} The indices in $\tilde{\pi}$ are satisfy condition (iii) in Definition \ref{def:freeness}, i.e, the adjacent indices are not equal.

We decompose $ \phi \left( \sum_{i=1}^n  A_i \right)^{2k}$ into two terms:

\begin{equation}
\phi \left( \sum_{i=1}^n  A_i \right)^{2k} =  \phi \sum_{\pi \in \Pi_1} \cdot +  \phi \sum_{\pi \in \Pi_2} \cdot : =  II_1 +  II_2,
\end{equation}
where the set of partitions $\Pi_1$ and $\Pi_2$ is defined as follows: Partition $\pi \in \Pi_1$ if and only if $\pi$ belongs to Case 1. And $\pi \in \Pi_2$ if and only if $\pi$ belongs to Case $2$.

By our assumption, i.e, freeness of $A_i,$ $II_2 =0.$ For the term $II_1,$ it is easy to see that $II_1$ is equal to the cardinality of the set $\Pi_1.$ Due to shrink process, $\pi \in \Pi_1$ only if there is even number of elements for every block. Those partitions with even elements in every block can be realized in following process: First choosing an arbitrary non-crossing pair partition, then combining some proper blocks to one block. Hence the number of $\pi \in \Pi_1$ is upper bounded by $C_k n^k.$ Thus
\begin{equation}
\phi \left( \sum_{i=1}^n  A_i \right)^{2k} \leq C_k n^k.
\end{equation}
Therefore under our assumption,
\begin{equation}
\left\| \sum_{i=1}^n A_i \right\| = \lim_{k \rightarrow \infty} \left( \phi \left( \sum_{i=1}^n  A_i \right)^{2k} \right)^{\frac{1}{2k}} \leq 2 \sqrt{n}.
\end{equation}

{\it Note}: For the local hidden state bound $C_n$ of steering inequality $S_n$ in equation \eqref{eq:LHS-bound}, the variables $\alpha_i$ do not make any effort to the whole derivation. Thus $C_n$ is also upper bounded by $2\sqrt{n}.$

%%%%%%%%%%%%%%%%%%%%%%%%%%%%%%%%%

\subsection{Fine-grained uncertainty for random observables}\label{sub:random-steering}

In fact, the statement is a corollary of the work of B. Collins and C. Male \cite{BM2013}. Here we restate their result as following:
Let $A_i = U_i D U_i^\dagger,$ then there exist $C^*$-probability space $(\mathcal{A}, \phi)$ and Haar unitaries $u_1, \ldots, u_n$ which are freely independent of element $d\in \mathcal{A},$ such that
\begin{equation}
\lim_{N \to \infty} \left\| \sum_{i=1}^n A_i \right\| = \left\| \sum_{i=1}^n u_i d u_i^* \right\|, \;\; \text{almost surely}.
\end{equation}

Denote $a_i = u_i d u_i^*,$ it is easy to see that $a_1, \ldots, a_n$ are freely independent in $(\mathcal{A}, \phi).$ Hence due to a similar argument in Appendix C, we have $\left\| \sum_{i=1}^n u_i d u_i^* \right\| \leq 2 \sqrt{n}.$ Therefore we have following corollary:

\begin{corollary}
Let $A_i = U_i D U_i^\dagger, i=1,\ldots, n,$ $U_i$ are independent random matrices in $U (N).$ Then we have
\begin{equation}
\lim_{N \to \infty}  \left\| \sum_{i=1}^n A_i \right\|  \leq 2 \sqrt{n}, \;\; \text{almost surely.}
\end{equation}
\end{corollary}

For the special case $n=2,$ we have following corollary:
\begin{corollary}
Let $A_i = U_i D U_i^\dagger, i=1,2,$ $U_i$ are independent random matrices in $U (N).$ Then we have
\begin{equation}
\lim_{N \to \infty} \left\|\sum_{i=1}^2 A_i \right\| = 2, \;\; \text{almost surely}.
\end{equation}
\end{corollary}

{\it Proof.}
For all $k \in \mathbb{N}_0,$ then almost surely we have
\begin{equation}
\begin{split}
 \lim_{N \to \infty} \tr \left( A_1 +  A_2 \right)^{2k} & = \sum_{j=0}^k \binom{k}{j} 2^{k-j} \sum_{l=0}^j \binom{j}{l} \lim_{N \to \infty} \tr \left( A_1 A_2\right)^{2l-j}\\
& = \sum_{j=0, \; \text{even}}^k \binom{k}{j} 2^{k-j} \binom{j}{j/2}.
\end{split}
\end{equation}
Since $\lim_{k \to \infty} \left( \sum_{j=0, \; \text{even}}^k \binom{k}{j} 2^{k-j} \binom{j}{j/2} \right)^{\frac{1}{2k}} =2,$
then by the standard argument in this sequel, we have
\begin{equation}
\liminf_{N \to \infty}  \Big\|\sum_{i=1}^2 A_i \Big\| \geq 2 -\epsilon, \; \text{almost surely}.
\end{equation}
On the other hand, $\Big\|\sum_{i=1}^2 A_i \Big\| \leq 2$ is obvious.

%%%%%%%%%%%%%%%%%%%%%%%%%%%%%%%%%

\subsection{Maximally certain states for $\xi_{\vec{0}}$ and $\xi_{\vec{1}}$ in the case $n=2$}\label{sub:uncertainty}

Let $A_1= \lambda(g_1), A_2 = \lambda(g_2),$ where $g_1, g_2$ are generator of group $\ast_2 \mathbb{Z}_2.$ We need following notions.

\begin{definition}
A group G is amenable if there exists a state $\mu$ on $\ell_\infty(G)$ which is invariant
under the left translation action: i.e. for all $s\in G$ and $f \in \ell_\infty (G), \mu(s \cdot f ) = \mu (f).$
\end{definition}

\begin{definition}
Let G be a group, a F{\o}lner net (sequence) is a net of non-empty finite subsets $F_n \subset G$ such that $|F_n \cap g F_n|/|F_n| \to 1$ for all $g\in G.$ Where $g F_n$ denotes the subset $\{g h: h\in F_n\}.$
\end{definition}
For any $g \in G,$ there exists $N,$ such that for all $n \geq N,$ $g \in F_n.$ There are many characterizations of amenable groups.
\begin{proposition}\cite{BO2008}
Let G be a discrete group. The following are equivalent:
\begin{enumerate}[{\rm i)}]
\item G is amenable;
\item G has a F{\o}lner net (sequence);
\item For any finite subset $E\subset G$, we have $\frac{1}{|E|} \left\|\sum_{g\in E} \lambda(g) \right\| =1.$
\end{enumerate}
\end{proposition}

For instance, group $\ast_2\mathbb{Z}_2$ is amenable. Hence by above proposition, $\|\lambda(g_1)+ \lambda(g_2)\|=2.$ With above notions, we can formally define a state
\begin{equation}
\rho_n = \frac{1}{|F_n|} \sum_{g,h \in F_n} |g \rangle \langle h|,
\end{equation}
where $F_n$ is a F{\o}lner sequence of $G= \ast_2 \mathbb{Z}_2.$ Now we have:
\begin{equation}
\begin{split}
\lim_{n \rightarrow \infty} \Tr ((\lambda(g_1)+ \lambda(g_2)) \rho_n) & = \lim_{n \to \infty} \frac{1}{|F_n|}  \left( \sum_{g,h \in F_n} \langle h| g_1 g\rangle + \sum_{g,h \in F_n} \langle h| g_2 g \rangle \right)\\
& = \lim_{n\to \infty} \frac{1}{|F_n|} (|F_n \cap g_1 F_n|+ |F_n \cap g_2 F_n|) =2,
\end{split}
\end{equation}
where for the second equation we have used the property of F{\o}lner sequence. Thus in this approximate sense, the fine-grained uncertainty of
$A_1^0$ and $A_2^0$ is 1. Technically we can construct $\tilde{\rho}_n$ to approximate $\xi_{\vec{1}}.$ Firstly we define two subsets of $G = \ast_2{\mathbb{Z}_2}:$
\begin{equation}
G_1 = \{g_1, g_1g_2, g_1g_2g_1, \cdots\} \;\; \text{and} \;\; G_2 = \{g_2, g_2g_1, g_2g_1g_2, \cdots\}.
\end{equation}
In fact, $G_1$ (resp. $G_2$) is the subset of wards which begin with $g_1$ (resp. $g_2$). It is easy to see $G_1 \cup G_2 \cup \{e\} = G.$ Now we define a state:
\begin{equation}
|\tilde{\phi}_n \rangle = \frac{1}{\sqrt{|F_n|}} \sum_{g \in F_n} e^{i \theta_g} |g\rangle,
\end{equation}
where $F_n$ is still a F{\o}lner sequence of $G$ and
\begin{equation}\label{eq:theta}
\theta_g = \left \{ \begin{split}
 & \pi/2 & \quad \quad  g \in G_1,\\
 & -\pi/2 & \quad \quad g \in G_2,\\
 & 0 & \quad \quad g=e.
 \end{split} \right.\end{equation}
Let $\tilde{\rho}_n = |\tilde{\phi}_n\rangle\langle\tilde{\phi}_n|,$ then we have
\begin{equation}\label{eq:g}
\begin{split}
\lim_{n \to \infty} Tr ((\lambda(g_1) + \lambda(g_2))\tilde{\rho}_n) & = \lim_{n \to \infty} \frac{1}{|F_n|}  \sum_{g,h \in F_n} e^{i (\theta_g-\theta_h)} \left( \langle h| g_1 g\rangle + \langle h| g_2 g \rangle \right)\\
& = \lim_{n \to \infty} \frac{1}{|F_n|} \left(  \sum_{g \in F_n \cap g_1 F_n}  e^{i (\theta_g-\theta_{g_1g})} + \sum_{g \in F_n \cap g_2 F_n}  e^{i (\theta_g-\theta_{g_2g})} \right).
\end{split}
\end{equation}

For the first of term of right hand side, for large enough $n,$ we can say $e, g_1 \in F_n.$ Therefore $e, g_1 \in F_n \cap g_1 F_n$ for large enough $n.$ Then we have
\begin{equation}
\begin{split}
\frac{1}{|F_n|} \sum_{g \in F_n \cap g_1 F_n}  e^{i (\theta_g-\theta_{g_1g})} & = \frac{1}{|F_n|} \sum_{g \in F_n \cap g_1 F_n, g\neq \{e, g_1\}}  e^{i (\theta_g-\theta_{g_1g})}+ \frac{1}{|F_n|}   e^{i (\theta_e-\theta_{g_1e})}+ \frac{1}{|F_n|}  e^{i (\theta_{g_1}-\theta_{g_1g_1})}\\
& = \frac{1}{|F_n|} \sum_{g \in F_n \cap g_1 F_n, g\neq \{e, g_1\}} e^{i \pi} = -\frac{|F_n \cap g_1 F_n|-2}{|F_n|},
\end{split}
\end{equation}
where for the second equation we have used \eqref{eq:theta}. A similar argument is valid for the second term of right hand side of \eqref{eq:g}. Thus finally we have
\begin{equation}
\lim_{n \to \infty} \Tr ((\lambda(g_1) + \lambda(g_2))\tilde{\rho}_n) = -2.
\end{equation}

%%%%%%%%%%%%%%%%%%%%%%%%%%

\subsection{Quantum value of complex CHSH-Bell inequality}

In this appendix we will consider a Bell inequality which has a similar form to the CHSH-Bell inequality. The Bell operator is defined as follows:
\begin{equation}
\mathcal{B} = A_1 \otimes B_1 + A_1 \otimes B_2 + A_2 \otimes B_1 + \omega A_2 \otimes B_2,
\end{equation}
where $\omega = e^{\frac{2\pi i}{3}}.$ Here the observables are not dichotomic. Instead, there are three possible outcomes: $1, \omega, \omega^2.$ Thus $A_i, B_j$ are required to be unitaries and satisfy $A_i^3 = B_j^3 = \un$ for any $i,j =1,2.$ The classical value of this Bell functional is $\sqrt{7}.$

Now for the quantum value, we can assume $A_1 = B_1, A_2 = B_2$ and $A_1, A_2$ are freely independent in some $C^*$-probability space. Hence we have
\begin{equation}
\mathcal{B} \mathcal{B}^\dagger = 3 \un \otimes \un + (\un - \omega A) \otimes (\un- \omega A),
\end{equation}
where $A= A_1 A_2^\dagger + \omega A_2 A_1^\dagger.$ By binomial formula we have
\begin{equation}
\tr (\mathcal{B} \mathcal{B}^\dagger)^k = \sum_{j=0}^k \binom{k}{j} 3^{k-j} \left( \sum_{l=0, \text{l is even}}^{j} \binom{j}{l}\binom{l}{l/2}  \right)^2:= Q_k.
\end{equation}
On one hand, by  Stirling's formula, for even $l,$ $\binom{l}{l/2} \leq 2^l,$ thus
\begin{equation}
Q_k \leq  \sum_{j=0}^k \binom{k}{j} 3^{k-j} \left( \sum_{l=0}^{j} \binom{j}{l} 2^l  \right)^2 = 12^k.
\end{equation}
By Lemma \ref{lem:c-norm}, we have $\|\mathcal{B}\| \leq 2 \sqrt{3}.$ By a slightly adaption of the results in \cite{EKB2013}, where they provided a method to estimate the quantum value for given dichotomic Bell inequalities, we can conclude that $2\sqrt{3}$ is an upper bound for the quantum value of complex CHSH Bell inequality. In fact, this upper bound can be obtained by choosing:
\begin{equation}
A_1 = B_1 = \begin{pmatrix}  & 0 & 0 & 1 \\ & \omega^2 & 0 & 0  \\ & 0 & \omega & 0 \end{pmatrix}, \;  A_2 = B_2 = \begin{pmatrix}  & 0 & 0 & -\omega \\ & 0 & 1 & 0  \\ & \omega^2 & 0 & 0 \end{pmatrix}.
\end{equation}

On the other hand,
\begin{equation}
\begin{split}
Q_{2k} & = \sum_{j=0}^{k} \binom{2k}{2j} 3^{2k-2j} \left( \sum_{l=0}^j \binom{2j}{2l}\binom{2l}{l}  \right)^2 + \sum_{j=1}^{k} \binom{2k}{2j-1} 3^{2k-2j+1} \left( \sum_{l=0}^{j-1} \binom{2j-1}{2l}\binom{2l}{l}  \right)^2\\
& \approx \sum_{j=0}^{k} \binom{2k}{2j} 3^{2k-2j} \left( \sum_{l=0}^j \binom{2j}{2l} 2^{2l}  \right)^2 + \sum_{j=1}^{k} \binom{2k}{2j-1} 3^{2k-2j+1} \left( \sum_{l=0}^{j-1} \binom{2j-1}{2l} 2^{2l} \right)^2\\
& \gtrsim \sum_{j=0}^{k} \binom{2k}{2j} 3^{2k-2j} \left( \sum_{l=0}^j \binom{j}{l} 2^{2l}  \right)^2 + \sum_{j=1}^{k} \binom{2k}{2j-1} 3^{2k-2j+1} \left( \sum_{l=0}^{j-1} \binom{j-1}{l} 2^{2l}  \right)^2\\
& \approx \sum_{j=0}^{2k} \binom{2k}{j} 3^{2k-j} 5^{j} = 8^{2k}.
\end{split}
\end{equation}
Therefore $\|\mathcal{B}\| \geq 2\sqrt{2}> \sqrt{7}.$

This method is also promising for the famous MABK Bell inequalities \cite{WW2001,ZB2002}.

\end{document}